\documentclass[a4paper,12pt]{article}
\usepackage[utf8]{inputenc}
\usepackage[T1]{fontenc}
\pdfoutput=1
\usepackage[margin=1in]{geometry}
\usepackage{textcomp}
\usepackage{lmodern}
\usepackage{ae}
\usepackage[english]{babel}
\usepackage[intlimits]{amsmath}
\usepackage{amssymb}
\usepackage{bm}
\usepackage{amsthm}
\usepackage{floatrow}
\usepackage{exscale}
\usepackage{subcaption}
\usepackage{graphicx}
\usepackage{titlesec}
\usepackage{algorithm}
\usepackage{algorithmic}
\usepackage{listings}
\usepackage{mathtools}
\usepackage{array}
\usepackage{booktabs}
\usepackage{siunitx}
\usepackage{etoolbox}
\usepackage{floatrow}
\usepackage{comment}
\newfloatcommand{capbtabbox}{table}[][\FBwidth]
\newcommand{\Mod}[1]{\ (\mathrm{mod}\ #1)}

\DeclarePairedDelimiter\abs{\lvert}{\rvert}%
\DeclarePairedDelimiter\norm{\lVert}{\rVert}%
\DeclareMathOperator*{\argmin}{arg\,min}
\makeatletter
\let\oldabs\abs
\def\abs{\@ifstar{\oldabs}{\oldabs*}}
\let\oldnorm\norm
\def\norm{\@ifstar{\oldnorm}{\oldnorm*}}
\makeatother
\newcommand{\myVertTab}[2]{\begin{tabular}{@{}l@{}}#1\\ #2\end{tabular}}

\renewrobustcmd{\bfseries}{\fontseries{b}\selectfont}
\renewrobustcmd{\boldmath}{}
\newrobustcmd{\B}{\bfseries}

\usepackage{csquotes}

\usepackage[style=phys,citestyle=phys,giveninits=true,maxbibnames=6,uniquename=init,minbibnames=3]{biblatex}
\providecommand{\keywords}[1]
{
  \small	
  \textbf{\textit{Keywords---}} #1
}

\author{Matti Hanhela$^1$, Mikko Kettunen$^2$, Olli Gr\"ohn$^2$,\\ Marko Vauhkonen$^1$ and Ville Kolehmainen$^1$\\
\small$^1$ Department of Applied Physics, University of Eastern Finland\\
\small$^2$ A.I. Virtanen Institute for Molecular Sciences, University of Eastern Finland
}

\begin{document}
\title{Temporal Huber regularization for DCE-MRI}
\maketitle

\begin{abstract}
	\noindent Dynamic contrast-enhanced magnetic resonance imaging (DCE-MRI) is used to study microvascular structure and tissue perfusion. In DCE-MRI a bolus of gadolinium based contrast agent is injected into the blood stream and spatiotemporal changes induced by the contrast agent flow are estimated from a time series of MRI data. Sufficient time resolution can often only be obtained by using an imaging protocol which produces undersampled data for each image in the time series. This has led to the popularity of compressed sensing based image reconstruction approaches, where all the images in the time series are reconstructed simultaneously, and temporal coupling between the images is introduced into the problem by a sparsity promoting regularization functional. We propose the use of Huber penalty for temporal regularization in DCE-MRI, and compare it to total variation, total generalized variation and smoothness based temporal regularization models. We also study the effect of spatial regularization to the reconstruction and compare the reconstruction accuracy with different temporal resolutions due to varying undersampling. The approaches are tested using simulated and experimental radial golden angle DCE-MRI data from a rat brain specimen. The results indicate that Huber regularization produces similar reconstruction accuracy with the total variation based models, but the computation times are significantly faster. \\
\end{abstract}

\vfill
\noindent \large{This is a pre-print of an article published in the Journal of Mathematical Imaging and Vision. The final authenticated version is available online at: https://doi.org/10.1007/s10851-020-00985-2.}

\newpage
%%%%%%%%%%%%%%%%%%%%%%%%%%%%%%%%%%%%%%%%%%%%%%%%%%%%%%%%%%%%%%%%%
\section{Introduction}

Dynamic contrast-enhanced MRI (DCE-MRI) is an imaging method which is used to study microvascular structure and tissue perfusion. It is used widely, for example, in studies of antivascular drugs \cite{OCo+12,ZP10}, multiple sclerosis \cite{Cra+14,Gai+11}, blood-brain-barrier assessment after acute ischemic stroke \cite{Mer+17,Vil+17} and treatment monitoring in breast cancer \cite{Mar+04,Pic+05} and glioma \cite{Pil+15}. The operation principle in DCE-MRI is to inject a bolus of gadolinium based contrast agent into the blood stream and acquire a time series of MRI data with a suitable $T_1$-weighting to obtain a time series of 2D (or 3D) images which exhibit contrast changes induced by concentration changes of the contrast agent in the tissues.

High spatial and temporal resolution of the DCE image series is required for accurate analysis of the contrast agent dynamics. In many cases, sufficient time resolution can only be obtained by utilizing an imaging protocol which produces undersampled data for each image in the time series. However, this has the complication that reconstructing undersampled datasets with conventional MR image reconstruction methods such as regridding \cite{Jac+91} leads to noisy image series with poor spatial resolution.

Recently, the compressed sensing (CS) framework has led to significant advances in MRI with undersampled data. The theory of CS states that a signal that is sparse in some basis, which is also incoherent with the measurement basis, can be perfectly reconstructed from undersampled data with a high probability \cite{CRT06,Don06}. Compressed sensing based approaches have been developed for numerous applications of both static and dynamic MRI, see for example the review \cite{JFL15}.

Provided that the temporal resolution of the DCE image series is high enough, one can expect high redundancy in the image series in the sense that the image intensity changes between successive image frames are small and occur only in part of the image domain. In such cases, it can be highly beneficial to sample the k-space in a complementary manner between successive time steps, meaning that the undersampling scheme should not be identical among neighboring points in time but rather such that  complementary information is collected from successive time points.

One such complementary sampling scheme is the golden angle (GA) approach where the measurements are done in radial fashion and the angle between subsequent radial spokes, which is 111.25$^\circ$, is based on the golden ratio \cite{Win+07}. The GA measurements have the advantages that the measurements are inherently complementary (i.e., each new spoke has a different path in the k-space compared to the previous ones) and each measured spoke traverses through the central part of the k-space which contains large information content on the contrast changes in the images. In addition, the GA sampling allows setting the segment length (i.e., the number of measured spokes per image frame) and thus the temporal resolution of the image series in the image reconstruction stage after the measurements are done. CS has been successfully used in combination with the GA sampling approach in several publications, including \cite{Fen+16,Fen+14,Kim+16,Usm+13}.

The basic structure in the CS approaches to DCE is to reconstruct the whole time series of the images simultaneously using an appropriate joint reconstruction formulation where a temporal regularization functional is employed for coupling the data across the time series of images. The most popular approach has been to use total variation (TV) regularization to promote sparsity of the derivative of the pixel (or voxel) values in the time direction. Temporal TV regularization has also been complemented with simultaneous use of spatial TV regularization in \cite{Adl+09}, where both of the TV regularizers were used in the smoothed (differentiable) form \cite{AV94}. The performance of different sparsity promoting temporal regularization schemes without any spatial regularization has been compared in DCE-MRI of the breast with cartesian k-space sampling in \cite{Wan+17b}.

Though widely used, TV regularization in the time direction may not be an optimal choice for DCE-MRI since the tumour signals in DCE-MRI are smooth in the time direction. One of the well known properties of TV regularization is the staircasing effect \cite{Poc+10}, i.e. piecewise flat reconstruction of smooth signals, leading potentially to suboptimal accuracy in the reconstruction and analysis of the signals. 

The staircasing effect could be alleviated by using $L_2$-norm based temporal smoothness (TS) regularization \cite{AWD07} or total generalized variation (TGV) regularization which promotes piecewise linear solutions \cite{BKP10}. Both TS and TGV have been used in CS DCE-MRI \cite{AWD07,Wan+17a,Wan+17b}. In \cite{Ras+18}, TS was used in combination with a spatial regularization term which used infimal convolution of two total variation Bregman distances for incorporating structural a priori information from an anatomical prior image into the reconstruction of the dynamic image sequence.

Another possible method for alleviating the staircasing effect would be to use the Huber penalty function \cite{Hub64} on the temporal gradient to enforce temporal regularity. Huber-penalty is a piecewise defined function which promotes smooth changes for small signal variations, but also allows large outliers similar to the total variation regularizer.

The estimation of the pharmacokinetic parameters of tissues requires an accurate estimate of tracer concentration in the artery \cite{Tof+99}. Estimating the arterial input function (AIF) via population averaging can produce adequate results in some cases, however, using patient specific input function produces more accurate estimates of kinetic parameters \cite{PKB01}. The AIF should preferably be extracted from a signal of a nearby artery feeding the tissues of interest, but it has also been estimated from a suitable venous sinus or vein in cases when the feeding artery is not visible \cite{LV10}. Obtaining an accurate patient specific AIF needs accurate reconstruction of the vascular input signal. Therefore, the regularization methods used in reconstructing DCE-MR image series should, in addition to obtaining good reconstruction quality of the smoothly varying tumour tissue signal, also be able to reconstruct the more rapidly varying vascular signal accurately.

In this study, we consider reconstruction of dynamic DCE data using a joint reconstruction which is based on minimization of a functional that combines least squares data misfit of the dynamic data, spatial TV regularization for promoting sparsity of the image gradients and a temporal regularization term for promoting regularity in the time direction. Based on the joint reconstruction formulation, we evaluate four different temporal regularization schemes for DCE using the gold\-en angle measurement scheme and study the effect of segment length (i.e. the number of radial GA spokes used per image) on signal accuracy in tumour and vascular regions as well as the rest of the tissue. We propose a novel approach using Huber-penalty \cite{Hub64} for temporal regularization and compare it to the widely used temporal TV, as well as $L_2$-smoothness (TS) and TGV regularizers. The Huber approach is expected to provide on par reconstruction accuracy with the state of the art TV methods with a reduced computation time.

In addition, the significance of the spatial regularization is evaluated by also studying the usage of temporal TV with no spatial regularization. The evaluations are carried out using both simulated and experimental golden angle DCE data where both cases correspond to small animal imaging of a rat brain, but the methods are also applicable to clinical imaging. In the simulation study, the GA approach is combined with a concentric squares sampling which uses varying length radial spokes to cover also the corner areas of the k-space in the sampling trajectory, to reduce the effects of peripheral aliasing artefacts to the evaluation of the methods.

%%%%%%%%%%%%%%%%%%%%%%%%%%%%%%%%%%%%%%%%%%%%%%%%%%%%%%%%%%%%%%%%%
\section{Theory}

%%%%%%%%%%%%%%%%%%%%%%%%%%%%%%%%%%%%%%%%%%%%%%%%%%%%%%%%%%%%%%%%%
\subsection{Forward model}

The measurement model in MRI is of the form
\begin{equation}
m = \mathcal{F} u + e,
\end{equation}
where $m\in\mathbb{C}^M$ is the complex valued measurement vector, $\mathcal{F}$ is the discrete Fourier transform, $u\in\mathbb{C}^N$ is the complex valued image, where $N=n\cdot n$ is the number of pixels in the image, and $e$ models the measurement noise. In the case of a non-cartesian k-space sampling trajectory, the Fourier transform is often approximated with the non-uniform fast Fourier transform (NUFFT) operation \cite{FS03}.

When NUFFT is employed as the forward model, the measurement model can be written as
\begin{equation}
m = P\mathcal{F}S u + e,
\end{equation}
where $P$ is an interpolation matrix from cartesian k-space to the non-cartesian k-space trajectory and $S$ is a scaling matrix. Hereafter we denote $A\coloneqq P\mathcal{F}S$.

In addition, when considering dynamic MRI with a complementary k-space sampling, where different (undersampled) trajectories of the k-space are measured at different time points, the forward model changes depending on the time point. The forward model can then be written as
\begin{equation}
m^t = P^t\mathcal{F}S^t u^t+e^t =A^t u^t+e^t,
\end{equation}
where the superscript $t$ denotes the time index of the measurement and image series, and $m^t$ is the vector of k-space data for a single image in the time series.

%%%%%%%%%%%%%%%%%%%%%%%%%%%%%%%%%%%%%%%%%%%%%%%%%%%%%%%%%%%%%%%%%
\subsection{Joint reconstruction formulation of the dynamic inverse problem}

In this study, we consider a joint reconstruction formulation of the DCE-MRI problem and compare the performance of four different temporal regularization functionals for promoting temporal regularity of the image series. The joint reconstruction formulation is based on an $L_2$-data misfit functional for the measurement model, and a spatial total variation regularization functional for promoting sparsity of the gradient of each image \cite{ROF92}. Only the temporal regularization method changes between the TV, TS, Huber and TGV models. The joint reconstruction formulation used in all reconstructions is thus of the form
\begin{equation}
u^*=\argmin_{u=u^1,u^2,...,u^T} \sum_{t=1}^T\Big[ \norm{A^tu^t-m^t}_2^2+ \alpha\norm{\nabla_\mathrm{S}u^t}_1\Big] + \beta \mathcal{T}(u),
\label{eq:basic_model}
\end{equation}
where $T$ is the number of image frames in the problem, $\nabla_\mathrm{S}$ is the discrete spatial gradient operator, $\alpha$ and $\beta$ are regularization parameters for the spatial and temporal regularization terms respectively, and $\mathcal{T}$ is one of the temporal regularization functionals. Here, the isotropic form of the 2D spatial TV is used where the total variation functional for a complex valued image $u^t$ is defined as
\begin{equation} \label{spatTV}
\norm{\nabla_\mathrm{S}u^t}_1 =\sum_{k=1}^N \Big( (\operatorname{Re}(\mathrm{D}_\mathrm{x}^ku^t))^2+(\operatorname{Re}(\mathrm{D}_\mathrm{y}^ku^t))^2+\\
(\operatorname{Im}(\mathrm{D}_\mathrm{x}^ku^t))^2+(\operatorname{Im}(\mathrm{D}_\mathrm{y}^ku^t))^2\Big)^{1/2} ,
\end{equation}
where $\operatorname{Re}$ and $\operatorname{Im}$ denote taking the real and the imaginary part of the complex valued image, $k$ denotes the spatial index in the 2D images, and $D_x^k$ and $D_y^k$ are discrete forward differences in the horizontal and vertical image directions of the k'th pixel defined as
\begin{align}
\mathrm{D}_\mathrm{x}^ku^t &= 
\begin{cases*}
-u_k^t+u_{k+n}^t, & if $k \leq N-n$ \\
0, & otherwise
\end{cases*}
\\
\mathrm{D}_\mathrm{y}^ku^t &=
\begin{cases*}
-u_k^t +u_{k+1}^t, & if $k\Mod{n} \neq 0$\\
0, & otherwise,
\end{cases*}
\end{align}
where $n$ is the number of rows and columns in the image.

%%%%%%%%%%%%%%%%%%%%%%%%%%%%%%%%%%%%%%%%%%%%%%%%%%%%%%%%%%%%%%%%%
\subsection{Temporal regularization functionals}

%%%%%%%%%%%%%%%%%%%%%%%%%%%%%%%%%%%%%%%%%%%%%%%%%%%%%%%%%%%%%%%%%
\subsubsection{Total variation model}

In the total variation model \cite{ROF92}, the temporal regularization part of the functional is
\begin{equation}
\mathcal{T}(u) = \norm{\nabla_\mathrm{T}u}_1
= \sum_{t=1}^T\sum_{k=1}^N \sqrt{(\operatorname{Re} (\mathrm{D}_\mathrm{T}^tu_k))^2 +(\operatorname{Im} (\mathrm{D}_\mathrm{T}^tu_k))^2},
\end{equation}
where $u_k=u_k^1,...,u_k^T$ and $\mathrm{D}_\mathrm{T}^t$ is the discrete forward difference in the temporal direction of the t'th image defined as
\begin{equation}
\mathrm{D}_\mathrm{T}^tu_k=
\begin{cases*}
-u_k^t+u_k^{t+1}, & if $t \neq T$\\
0, & otherwise.
\end{cases*}
\end{equation}

The temporal total variation model promotes sparsity of the time derivative of the pixel signals, being highly feasible for reconstruction of piecewise regular signals which may exhibit large jumps. A similar regularization functional, but in the smoothed, differentiable form, was used in \cite{Adl+09} for multislice myocardial perfusion imaging. 

A well-known feature of the TV model is the so-called staircasing effect, where smooth signals are reconstructed as piecewise constant \cite{Poc+10}. This may potentially lead to suboptimal results in applications with smooth pixel signals.

To study the significance of spatial regularization in the joint reconstruction model \eqref{eq:basic_model},
we also consider the temporal TV model without spatial regularization, i.e. $\alpha$ in \eqref{eq:basic_model} is set to 0. We denote the temporal TV model without spatial regularization by TV-T.

%%%%%%%%%%%%%%%%%%%%%%%%%%%%%%%%%%%%%%%%%%%%%%%%%%%%%%%%%%%%%%%%%
\subsubsection{Smoothness model}

The temporal smoothness model promotes smooth, slow\-ly changing signals by using the squared $L_2$-norm of the time derivative for the temporal regularization, that is
\begin{equation}
	\mathcal{T}(u)=\norm{\nabla_\mathrm{T}u}_2^2=\sum_{t=1}^T\sum_{k=1}^N\Big[ \big( \operatorname{Re}\big( \mathrm{D}_\mathrm{T}^tu_k \big) \big)^2 + \big( \operatorname{Im}\big( \mathrm{D}_\mathrm{T}^tu_k \big) \big)^2 \Big].
\end{equation}
We refer to this as the temporal smoothness (TS) model.

The TS model generally reconstructs smooth signals well, but fast transient signal changes often get diminished. TS model has been used in \cite{AWD07} for radial DCE myocardial perfusion imaging, and temporal smoothness regularization was compared with temporal TV regularization in the same application in \cite{AD08}.

%%%%%%%%%%%%%%%%%%%%%%%%%%%%%%%%%%%%%%%%%%%%%%%%%%%%%%%%%%%%%%%%%
\subsubsection{Huber model}

In the Huber model, the Huber penalty function \cite{Hub64}
\begin{equation}
	H_\gamma(v)=\begin{cases}
	\frac{\abs{v}^2}{2\gamma},\ &\abs{v}\leq \gamma\\
	\abs{v} - \frac{\gamma}{2},\ &\abs{v}>\gamma
	\end{cases}
\end{equation}
is used for the regularization of the time derivative. The Huber penalty function has quadratic growth near origin and linear growth further from origin. The transition point from quadratic to linear is controlled by the Huber parameter $\gamma$. When the parameter $\gamma$ is small, Huber regularization is close to TV regularization and when the parameter is large, Huber regularization is related to smoothness regularization.

The discrete temporal Huber regularization functional is of the form
\begin{equation}\label{HuberT}
    \begin{aligned}
    	\mathcal{T}(u) &= \sum_{t=1}^T \sum_{k=1}^NH_\gamma(\nabla_\mathrm{T}^tu_k) \\&= \sum_{(k,t)\in G_1}\left( \frac{(\operatorname{Re}(\mathrm{D}_\mathrm{T}^tu_k))^2 + (\operatorname{Im}(\mathrm{D}_\mathrm{T}^tu_k))^2}{2\gamma} \right) \\&+ \sum_{(k,t)\in G_2}\left( \sqrt{(\operatorname{Re}(\mathrm{D}_\mathrm{T}^tu_k))^2 + (\operatorname{Im}(\mathrm{D}_\mathrm{T}^tu_k))^2} - \frac{\gamma}{2}\right),
	\end{aligned}
\end{equation}
where $G_1=\{k\in{1,..,N},t\in{1,..,T}\mid \abs{\nabla_\mathrm{T}^t u_k} \leq \gamma \}$ and $G_2=\{k\in{1,..,N},t\in{1,..,T}\mid \abs{\nabla_\mathrm{T}^t u_k} > \gamma \}$.

The Huber model is expected to produce smooth signals for small variations in the signal while also allowing a few jumps (discontinuities) in the signal. The Huber model parameter $\gamma$ also has a physical interpretation; it defines the threshold for a signal change that is assumed to be a discontinuous jump.

%%%%%%%%%%%%%%%%%%%%%%%%%%%%%%%%%%%%%%%%%%%%%%%%%%%%%%%%%%%%%%%%%
\subsubsection{Total generalized variation model}

The total generalized variation model \cite{BKP10} is a total variation model, which is generalized to higher order differences. Here we use the second-order total generalized variation, which in the discrete 1-dimensional form is of the form
\begin{equation}
	\mathcal{T}(u) = \mathrm{TGV}_\gamma^2(u)=\min_v\norm{\nabla_\mathrm{T}u-v}_1+\gamma \norm{\nabla_\mathrm{T}v}_1.
\end{equation}
This functional balances between minimizing the first-order and second-order differences of the signal. The difference with TV-regularization is the most clear in smooth regions where piecewise linear solutions are favored over the piecewise constant solutions of TV.

TGV was first used in MRI as a spatial prior in \cite{Kno+11}. TGV has also been used in MRI as a temporal prior in \cite{Wan+17a}, where different temporal priors were compared in cartesian DCE-MRI of the breast.

%%%%%%%%%%%%%%%%%%%%%%%%%%%%%%%%%%%%%%%%%%%%%%%%%%%%%%%%%%%%%%%%%
\section{Methods}
The joint reconstructions with different regularization schemes are evaluated using simulated golden angle DCE-MRI data from a rat brain phantom and with experimental golden angle DCE-MRI data from a rat glioma model.

%%%%%%%%%%%%%%%%%%%%%%%%%%%%%%%%%%%%%%%%%%%%%%%%%%%%%%%%%%%%%%%%%
\subsection{Simulation}
\begin{figure}
	\centering
	\includegraphics[width=.8\columnwidth]{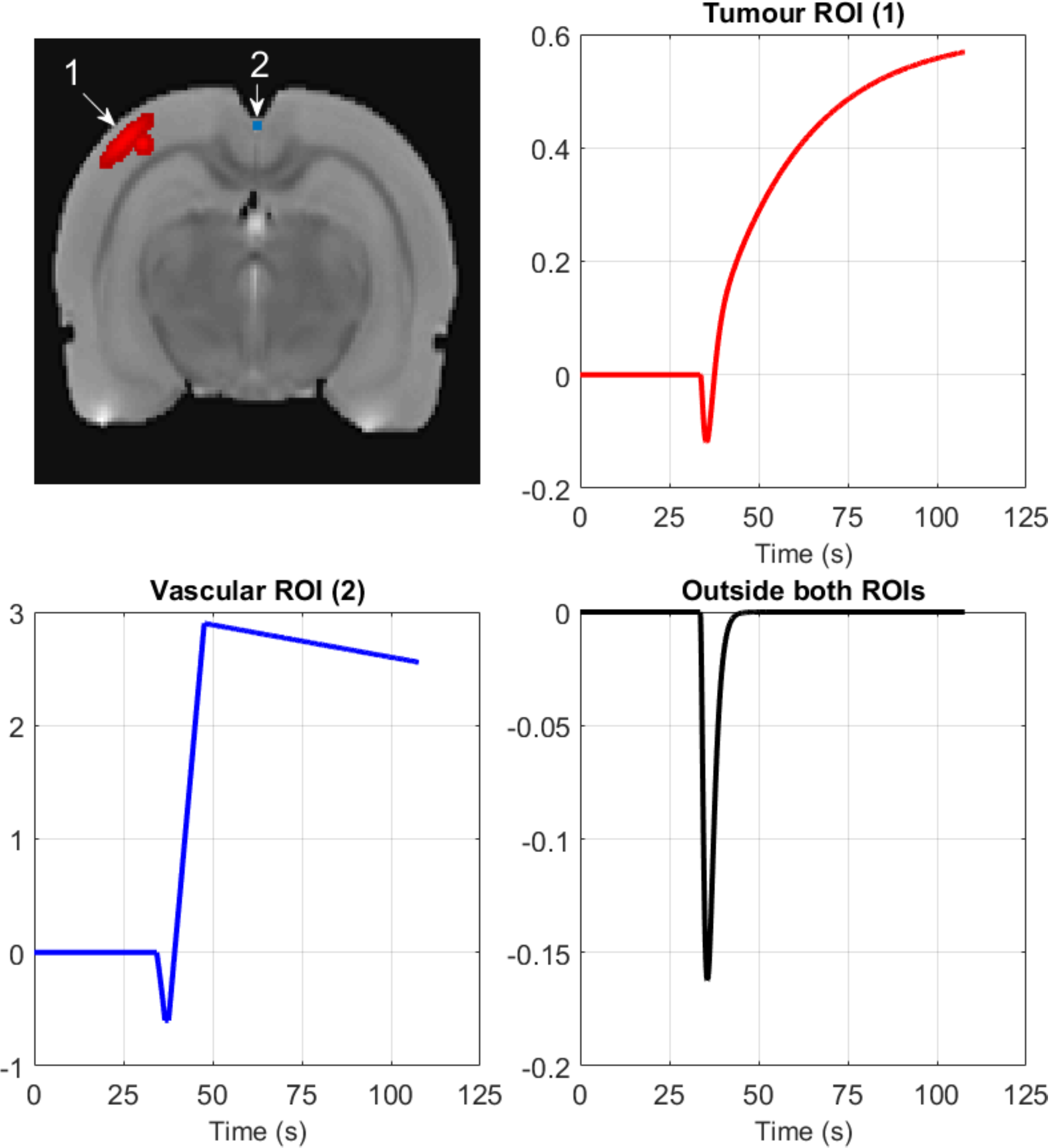}
	\caption{Simulated signals in different ROIs. Top left: The simulated image with tumour ROI marked in red and labelled '1', and vascular ROI marked in blue and labelled '2'. Top right: Simulated tumour ROI signal. Bottom left: Simulated vascular ROI signal. Bottom right: Simulated signal in tissue outside both ROIs. The vertical axis in the three figures is the multiplier for the signal added to the base signal and not the added signal itself.} 
	\label{fig:simu}
\end{figure}

A simulated test case modelling DCE-MRI of a tumour in rat brain was created. The rat brain phantom is based on the rat brain atlas in \cite{Val+11}, and scaled to a size of 128x128. The rat brain image was divided into three subdomains of different signal behaviour identified in the in vivo measurements: simulated tumour (region highlighted with red and labelled '1' in Fig.~\ref{fig:simu}), vascular region (highlighted with blue and labelled '2' in Fig.~\ref{fig:simu}) and the rest of the brain. The vascular signal region corresponds to the location of the superior sagittal sinus which can be used for estimating the AIF in DCE-MRI of the brain \cite{LV10}.

A time series of 2800 ground truth images was simulated by multiplying the signal of each pixel with the template of the corresponding region and adding that to the baseline value of the pixels. The tumour signal templates were based on an experimental DCE-MRI measurement, which is described in Sect.~\ref{ssec:invivomeas}, where the three different ROIs were identified. Fig.~\ref{fig:simu} shows the signal templates for each of the different tissue regions.

One spoke of k-space data was simulated for each of the simulated images, leading to a dynamic experiment with 2800 spokes of k-space data. The time scale of the simulation was set to be similar to the in vivo measurements where the repetition time of the GA measurements was 38.5ms. Gaussian complex noise at 5\% of the mean of the absolute values of the signal was added to the simulated k-space signal.

\begin{figure}
\captionsetup[sub]{justification=centering}
    \begin{subfigure}[b]{0.24\columnwidth}
    	\centering
    	\includegraphics[width=.9\textwidth]{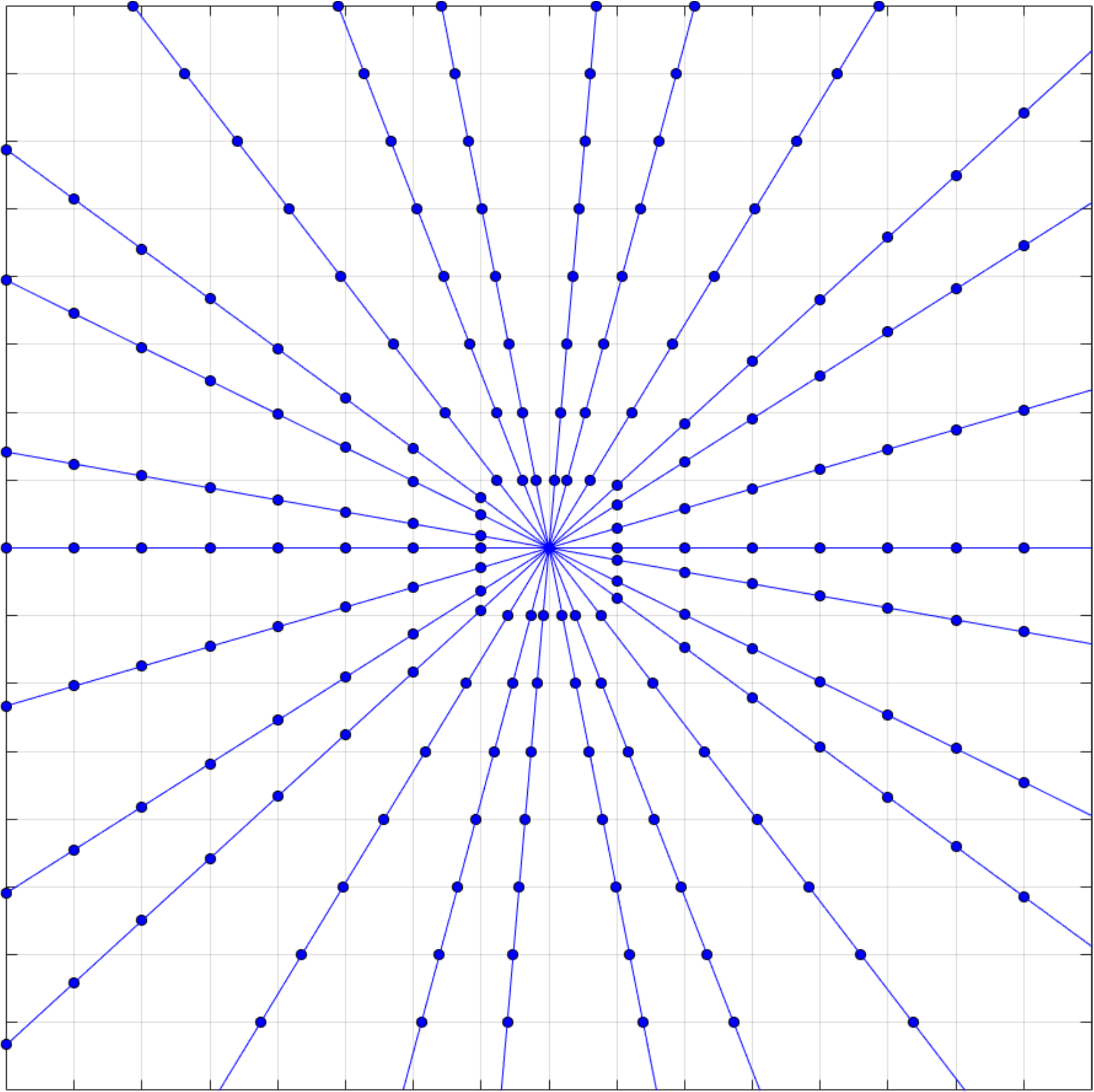}
    	\caption{}
    	\label{fig:sampling}
    \end{subfigure}%
    \begin{subfigure}[b]{0.76\columnwidth}
    	\centering
    	\includegraphics[width=.9\textwidth]{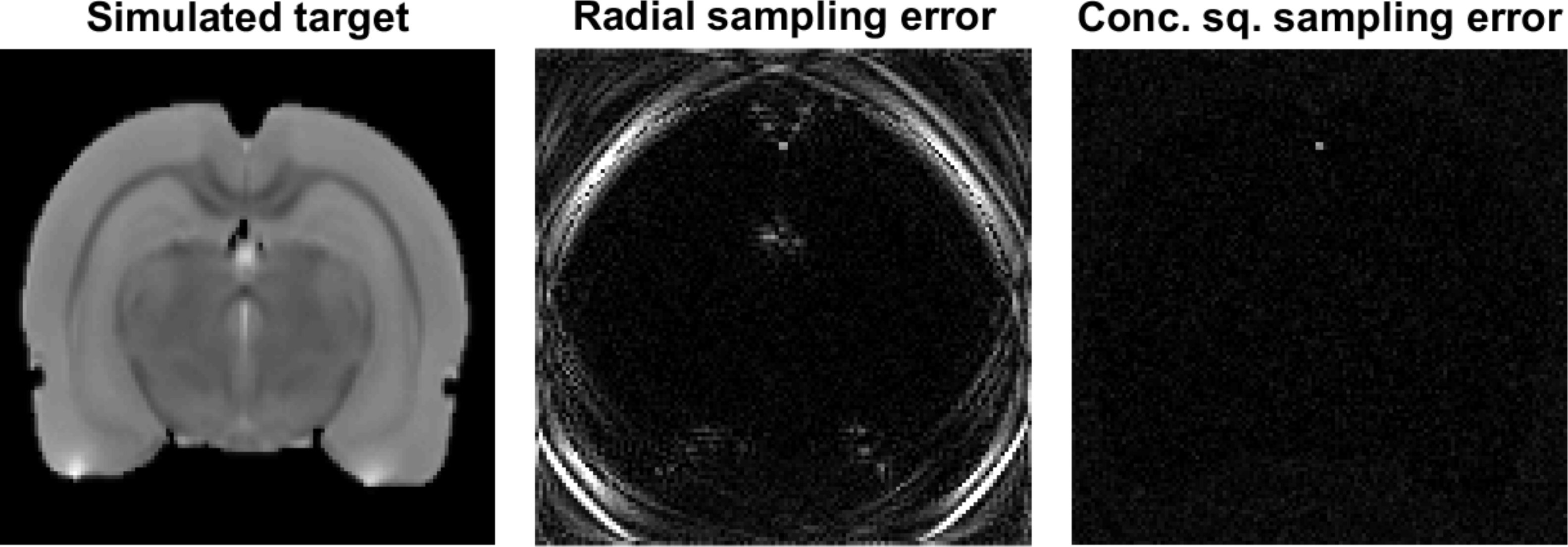}
    	\caption{}
    	\label{fig:rad_vs_rect}
	\end{subfigure}
	\caption{a) Center of the concentric squares sampling grid with the dots marking the sampled points and lines indicating the set of points sampled after a single echo. The measurement points form concentric squares instead of concentric circles as in a conventional radial measurement. b) The simulated target (Left) and reconstruction error images of inverse NUFFT reconstructions from 2000 GA spokes of conventional radially sampled data (Middle) and concentric squares sampled data (Right). The error images have the same color scale.}
\end{figure}

The simulated test case was carried out using a k-space trajectory which combines the golden angle and the concentric squares sampling strategies. The k-space trajectory of this sampling is illustrated for a few consecutive spokes in Fig.~\ref{fig:sampling}. This sampling strategy is similar to the linogram method \cite{EH87} developed for computed tomography imaging, but the lines in linogram sampling are equidistant in $\tan\theta$, whereas here the angles were chosen according to the golden angle method \cite{Win+07}.

Unlike the conventional radial sampling pattern with spokes of equal length, the concentric squares sampling strategy also covers the corners of the k-space. The sampling pattern therefore also collects information of the high frequencies in the corners of the k-space, leading to reduction of artefacts caused by the lack of sampling in the corners. This sampling pattern thus allows better comparison of the different methods compared to conventional radial GA sampling. This artefact reduction is demonstrated in Fig.~\ref{fig:rad_vs_rect}, where error images of inverse NUFFT reconstructions from 2000 GA spokes of conventional radial sampling and concentric squares sampling are shown. With 2000 spokes of data, the inverse NUFFT reconstruction of conventional radially sampled data has significant artefacts in the peripheral regions of the image domain, whereas the reconstruction of concentric squares sampled data does not.

In the NUFFT implementation, the measurements were interpolated into a twice oversampled cartesian grid with min-max Kaiser-Bessel interpolation with a neighbourhood size of 4 \cite{FS03}. Compared to a conventional radial sampling pattern, the interpolation distances of the radial concentric squares sampling strategy are shorter to the cartesian grid, resulting in smaller interpolation error. If the measurement angles are equi\-distant in slope rather than angle, the pseudo-polar Fourier transform can also be applied \cite{ACD+06}. The pseudo-polar FT has been used in CS MRI in \cite{YLL+16}.

%%%%%%%%%%%%%%%%%%%%%%%%%%%%%%%%%%%%%%%%%%%%%%%%%%%%%%%%%%%%%%%%%
\subsection{In vivo measurements from a rat}\label{ssec:invivomeas}

%%%%%%%%%%%%%%%%%%%%%%%%%%%%%%%%%%%%%%%%%%%%%%%%%%%%%%%%%%%%%%%%%
\subsubsection{Animal preparation}

All animal experiments were approved by the Animal Health Welfare and Ethics Committee of University of Eastern Finland. 1x$10^6$ C6 (ECACC 92090409) rat glioma cells (Sigma) were implanted into the brain of a 200 g female Wistar rat under ketamin/medetomidine hydrochloride anesthesia. Tumor imaging was performed 10 days post-implantation. During the experiments, the animal was anesthetized with isoflurane (5 \% induction, 1-2 \% upkeep) and kept in fixed position in a holder which was inserted into the magnet. A needle was placed into the tail vein of the animal for the injection of the contrast agent.

%%%%%%%%%%%%%%%%%%%%%%%%%%%%%%%%%%%%%%%%%%%%%%%%%%%%%%%%%%%%%%%%%
\subsubsection{Acquisition of the data}
The experimental small animal data were collected using a 9.4 T horizontal magnet interfaced to Agilent imaging console and a volume coil transmit/quadrature surface coil receive pair (Rapid Biomed, Rimpar, Germany). The data were collected with conventional radial golden angle sampling using a gradient-echo based radial pulse sequence with repetition time 38.5 ms, echo time 9 ms, flip angle 30 degrees, field-of view 32 mm x 32 mm, slice thickness 1.5 mm, number of points in each spoke 128. 610 spokes were collected in sequential order, after which the next spoke would differ less than 0.1 degrees from zero, so to simplify the experimental sequence, the cycle of 610 spokes was repeated for 25 times, leading to an overall measurement sequence of 15250 spokes of data for a total measurement duration of nearly 10 minutes.

In the computations, 7310 spokes of data were used for evaluation of the different models to speed up the computations, starting from the beginning of the measurements. Measurement time for a full cycle of 610 spokes was $610 \cdot 38.5$ ms $= 23.46$ s. Gadovist (1 mmol/kg) was injected i.v. one minute after the beginning of the dynamic scan over a period of 3 s.

\begin{figure}
	\centering
	\includegraphics[width=.8\columnwidth]{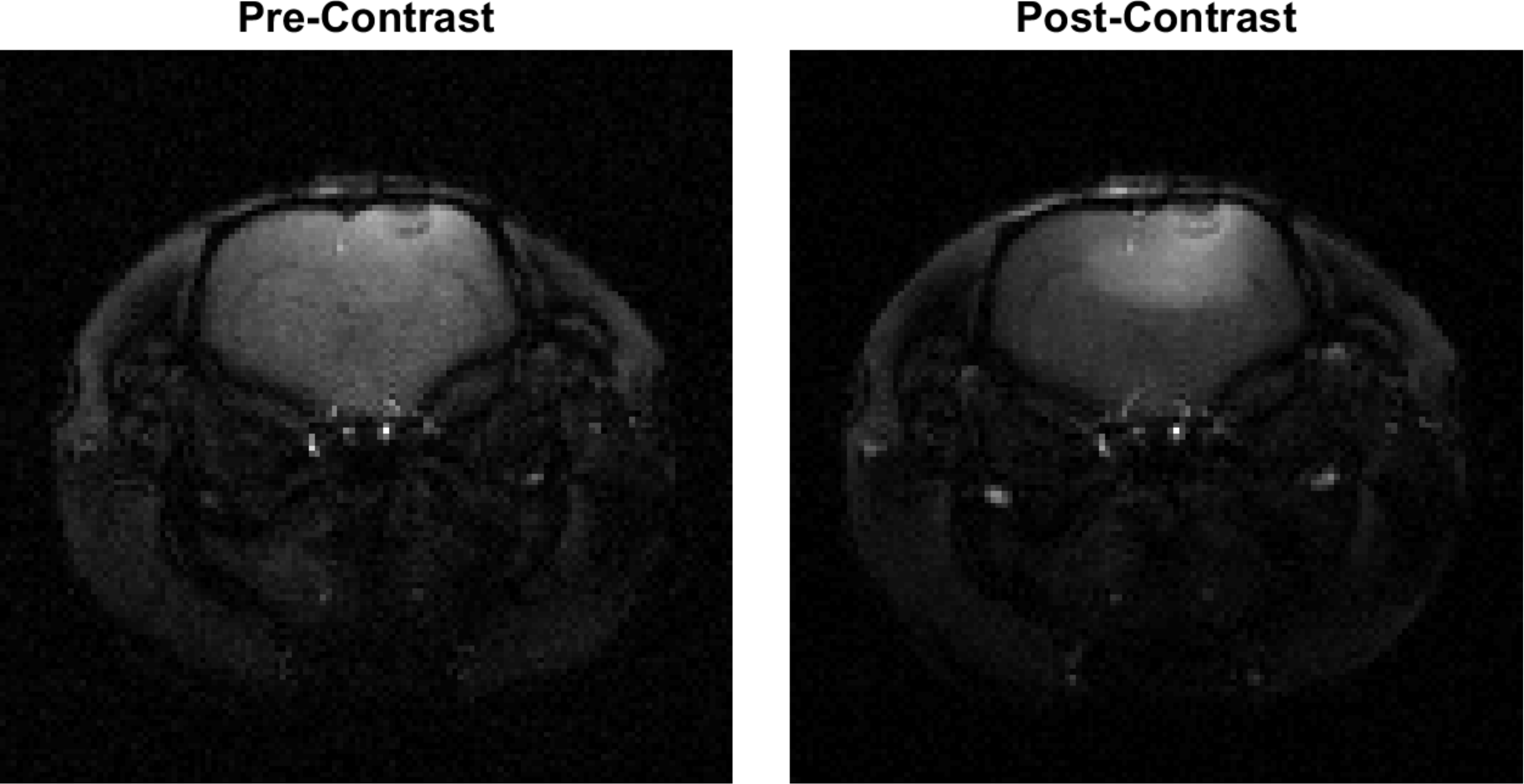}
	\caption{Cartesian gradient-echo pulse sequence full data IFFT reconstructions from before and after contrast injection used as reference. The two images have different adjusted color scales for visualization purposes.}
	\label{fig:sems_ref}
\end{figure}

Anatomical images were acquired from the same slice before and after the dynamical experiment using a gradient-echo pulse sequence with otherwise similar parameters as in the dynamic sequence but using a Cartesian sampling of 128x128 points of k-space data. The full data anatomical images from before and after the experiment are shown as reference for the dynamical reconstructions with undersampled data in Fig.~\ref{fig:sems_ref}.

The dynamical experiment also served as a basis for creating the signal templates shown in Fig.~\ref{fig:simu} for the simulated test case. For the simulation, three regions were identified from the in vivo reconstructions: vascular region (superior sagittal sinus), glioma region, and the rest of the brain tissue.

%%%%%%%%%%%%%%%%%%%%%%%%%%%%%%%%%%%%%%%%%%%%%%%%%%%%%%%%%%%%%%%%%
\subsection{Computation}
The Chambolle-Pock primal-dual algorithm \cite{CP11,SJP12} was used for all image reconstructions. The ratio of the primal and dual step sizes was varied according to the regularization coefficient and method, such that the primal step size was smaller for larger regularization parameters and the smooth Huber-regularization and TS-regularization had less variation in the step lengths. Usage of asymmetrical step sizes in the algorithm has been shown to lead to faster convergence in some cases in both linear \cite{PC11} and non-linear \cite{Val14} problems. The operator norm of the forward problem was calculated with the power method and the operator was scaled to have a norm of $\norm{A}=\sqrt{12}$ to be on the same order of magnitude as the difference operators which were used in the computation of the image gradients.

%%%%%%%%%%%%%%%%%%%%%%%%%%%%%%%%%%%%%%%%%%%%%%%%%%%%%%%%%%%%%%%%%
\subsubsection{Error metric}
Root mean square error (RMSE) values were first calculated for the three regions (tumour region, vascular region, rest of the image) separately after the reconstructed signals of each pixel were linearly interpolated in the temporal direction to match with the temporal resolution of segment length of one. Thus, after the interpolation all signals reconstructed with different segment lengths had the same number of time points as the series of the ground truth images enabling the comparison of reconstructions with different segment lengths. Specifically, the RMSEs were calculated by
\begin{equation}
    \mathrm{RMSE}_{\mathrm{ROIn}}=\sum_{k\in \Omega_{\mathrm{ROIn}}}\sqrt{\frac{\sum_{t=1}^T\left( \left( \abs{u_{k}^{t,\mathrm{interp}}}-\abs{I_k^t}\right)^2\right)}{T}},
\end{equation}
where $\Omega_{\mathrm{ROIn}}$ denotes the pixels in the n'th ROI, $T$ is the number of simulated time frames and measurement spokes, $u^\mathrm{interp}$ is the time-interpolated reconstructed pixel signal and $I$ is the ground truth image. 

After the separate ROI RMSEs were computed, a joint RMSE was computed by taking the norm of the separate RMSEs
\begin{equation}
    \mathrm{RMSE}_{\mathrm{joint}} = \sqrt{\mathrm{RMSE}_{\mathrm{ROI1}}^2 + \mathrm{RMSE}_{\mathrm{ROI2}}^2 +\mathrm{RMSE}_{\mathrm{ROI3}}^2}.
\end{equation}
This was done to weigh the small ROIs appropriately in the error metric that is used for performance comparisons.

%%%%%%%%%%%%%%%%%%%%%%%%%%%%%%%%%%%%%%%%%%%%%%%%%%%%%%%%%%%%%%%%%
\section{Results}

%%%%%%%%%%%%%%%%%%%%%%%%%%%%%%%%%%%%%%%%%%%%%%%%%%%%%%%%%%%%%%%%%
\subsection{Simulation}

Reconstructions with varying segment lengths and different temporal regularization methods were calculated for a wide range of the temporal regularization parameter. The tested segment lengths were chosen to be 8, 13, 21, 34, 55 and 89. These segment lengths are Fibonacci numbers, which provide optimal measurement profile distribution when golden angle measurements are used \cite{Win+07}.

The spatial regularization parameter was constant at $\alpha=0.001$ in all reconstructions. This level of spatial regularization was found to provide accurate reconstructions with all segment lengths and temporal regularization models. For TGV, the ratio of the first and second order terms was set to $\gamma=\sqrt{2}$ as in \cite{BH15}. For the Huber model, the Huber parameter was set to $\gamma=0.001$. This value corresponds to approximately 92\%-96\% of the simulated pixel changes between true images at intervals matching the varying segment lengths being under the threshold.

The optimal temporal regularization parameter $\beta$ for each regularization method and segment length was selected to be the one yielding the lowest joint RMSE. Table \ref{tab:opt_params} shows the optimal regularization parameters for the different methods with respect to segment length. In all cases optimal temporal regularization parameter increases monotonically as segment length decreases. This behavior is expected, since shorter segment length means less data per image and therefore the reconstructions require stronger regularization. The Huber method has the smallest change in the regularization parameter when segment length changes, whereas the optimal parameter for the TS model changes by multiple orders of magnitude when segment length is changed.

\begin{figure}[hbtp]
	\centering
	\begin{subfigure}[b]{0.45\columnwidth}
		\centering
		\includegraphics[width=.95\textwidth]{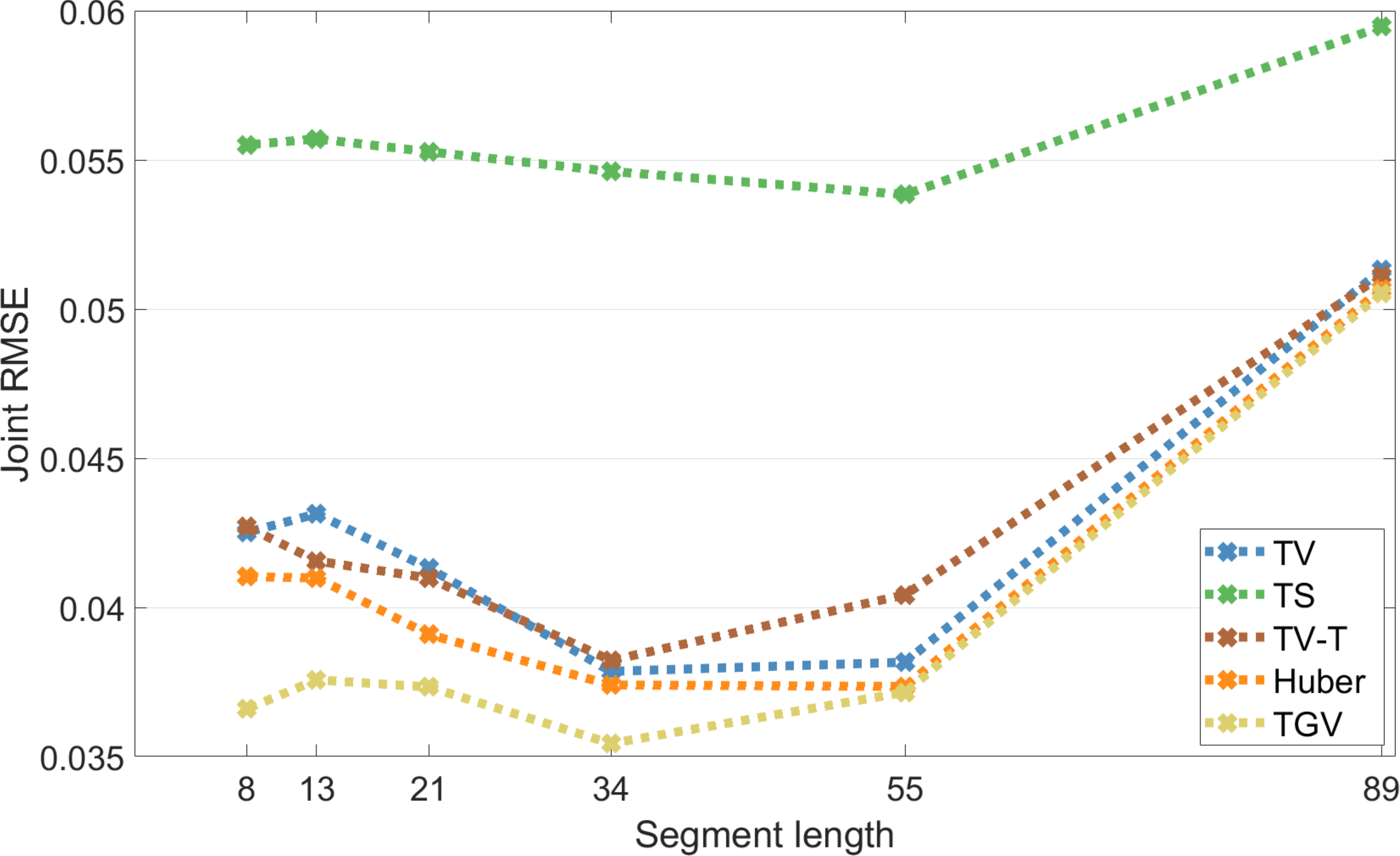}
		\caption{Joint RMSE}
		\label{fig:jrmse}
	\end{subfigure}%
	\begin{subfigure}[b]{0.45\columnwidth}
		\centering
		\includegraphics[width=.95\textwidth]{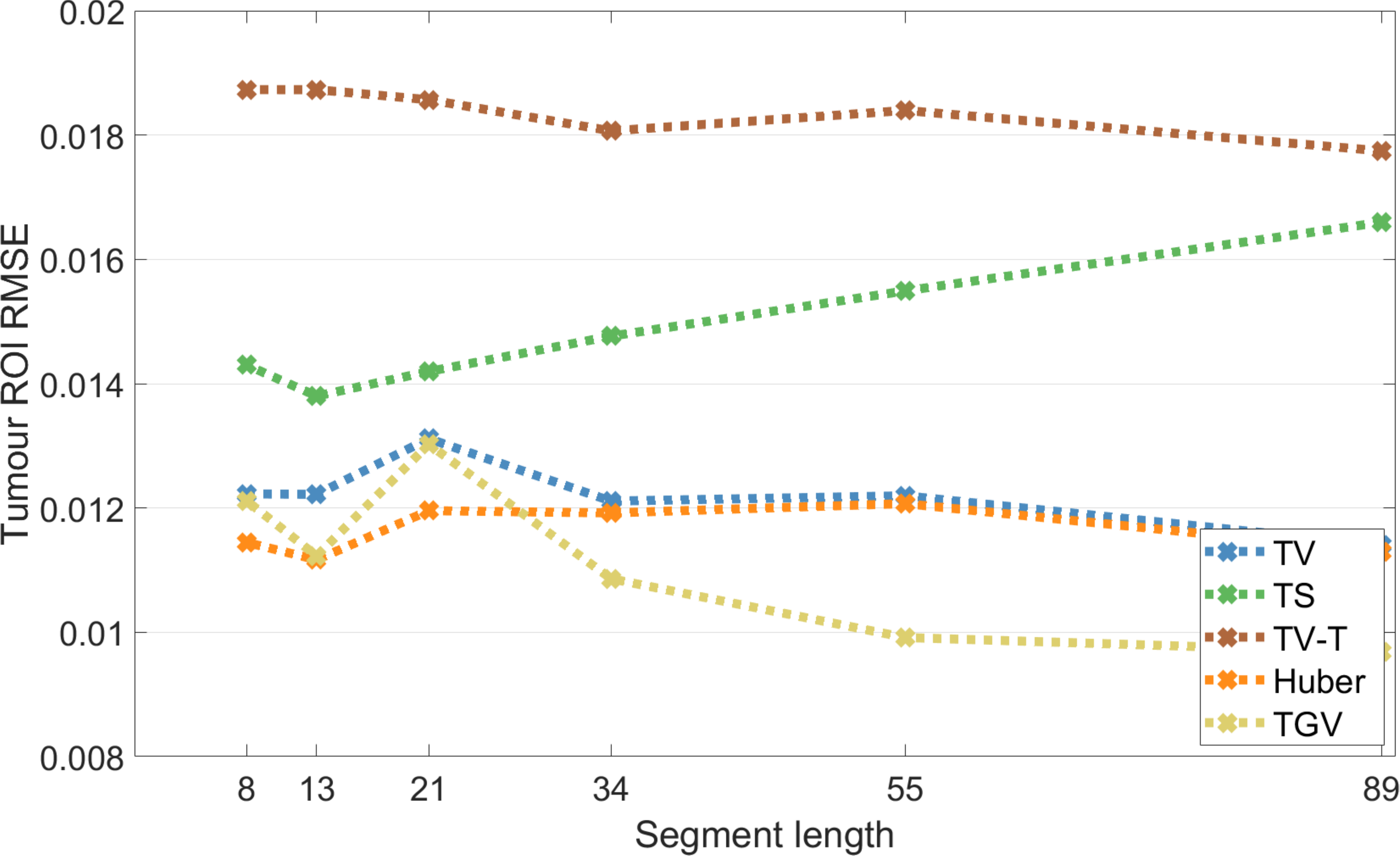}
		\caption{Tumour ROI RMSE}
		\label{fig:roi1_rmse}
	\end{subfigure}
	\newline
	\begin{subfigure}[b]{0.45\columnwidth}
		\centering
		\includegraphics[width=.95\textwidth]{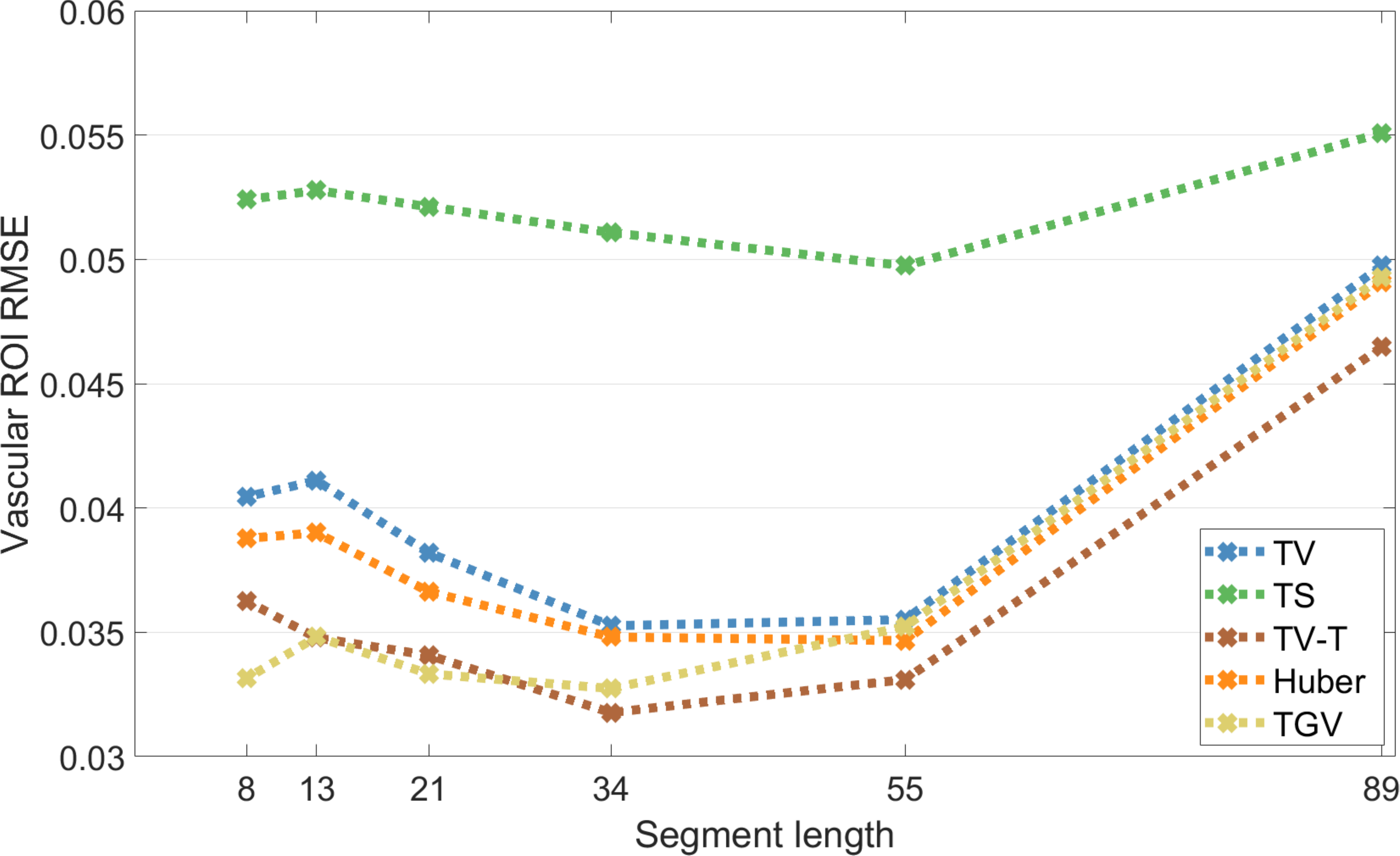}
		\caption{Vascular ROI RMSE}
		\label{fig:roi2_rmse}
	\end{subfigure}%
	\begin{subfigure}[b]{0.45\columnwidth}
		\centering
		\includegraphics[width=.95\textwidth]{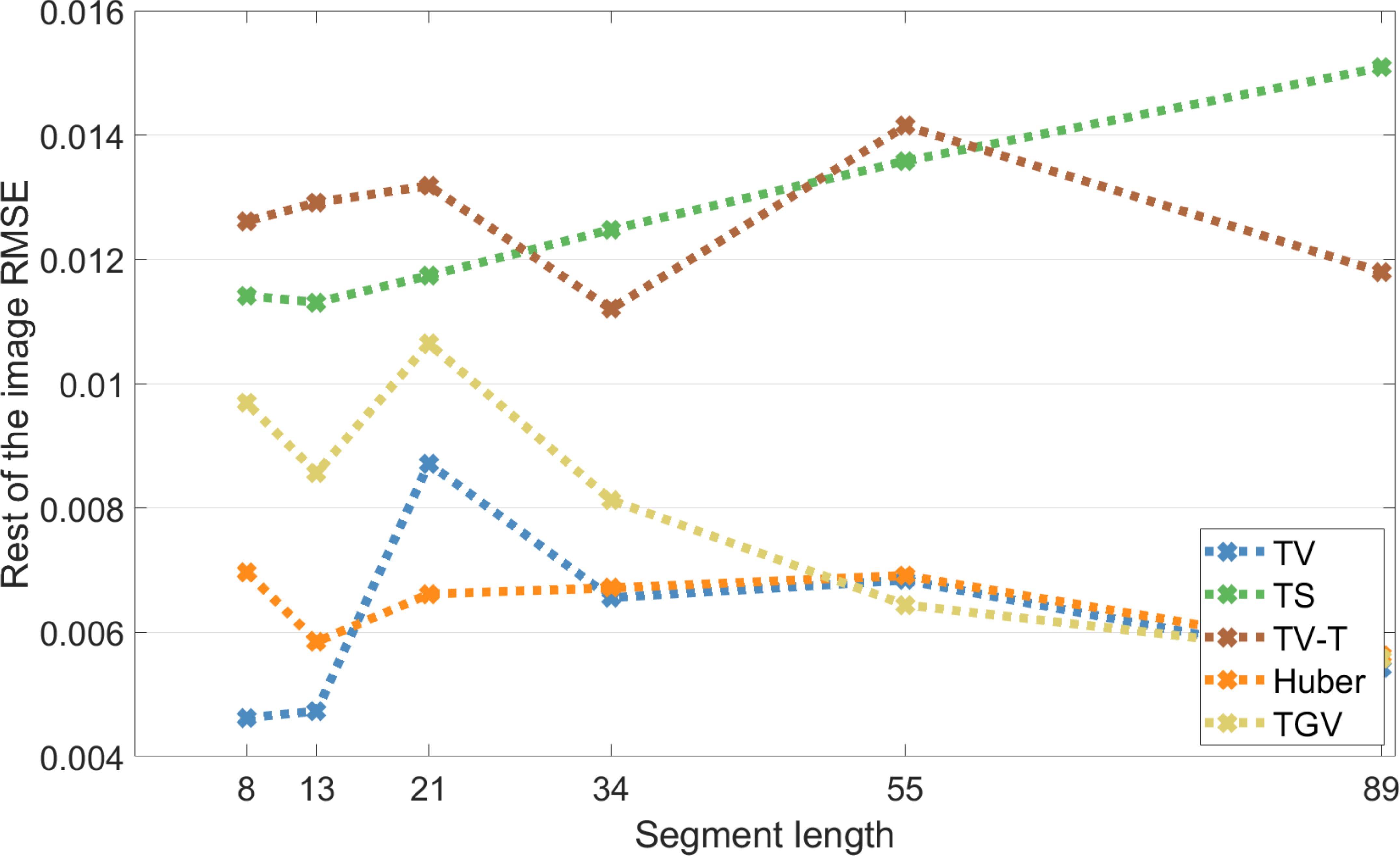}
		\caption{Rest of the image RMSE}
		\label{fig:roi3_rmse}
	\end{subfigure}
	\caption{a) Joint RMSE and b), c), d) RMSEs of the three different regions at different segment lengths with the optimal temporal regularization parameters for each segment length selected according to lowest joint RMSE.}
	\label{fig:roi_rmses}
\end{figure}

Table \ref{tab:rmse_rois_res} and Fig.~\ref{fig:jrmse} show the joint RMSE with different segment lengths and temporal regularization methods with the optimal temporal regularization parameters. Segment length of 34 is optimal for all but TS and Huber models, for which segment length of 55 is slightly better. For TV, the reconstruction accuracy with segment lengths of 34 and 55 are also very close. TGV produces the most accurate reconstructions for all segment lengths. TS performs the worst here with all segment lengths, which is due to it's poor performance in the vascular ROI accuracy as is evident from Fig.~\ref{fig:roi2_rmse}.

While the TV-T model without spatial regularization performs well in reconstructing the vascular ROI as seen in Fig.~\ref{fig:roi2_rmse}, it does considerably worse in reconstructing the tumour ROI and the rest of the image as seen in Figs.~\ref{fig:roi1_rmse} and \ref{fig:roi3_rmse}.

\begin{figure}[hbtp]
    \centering
    \includegraphics[width=.9\columnwidth]{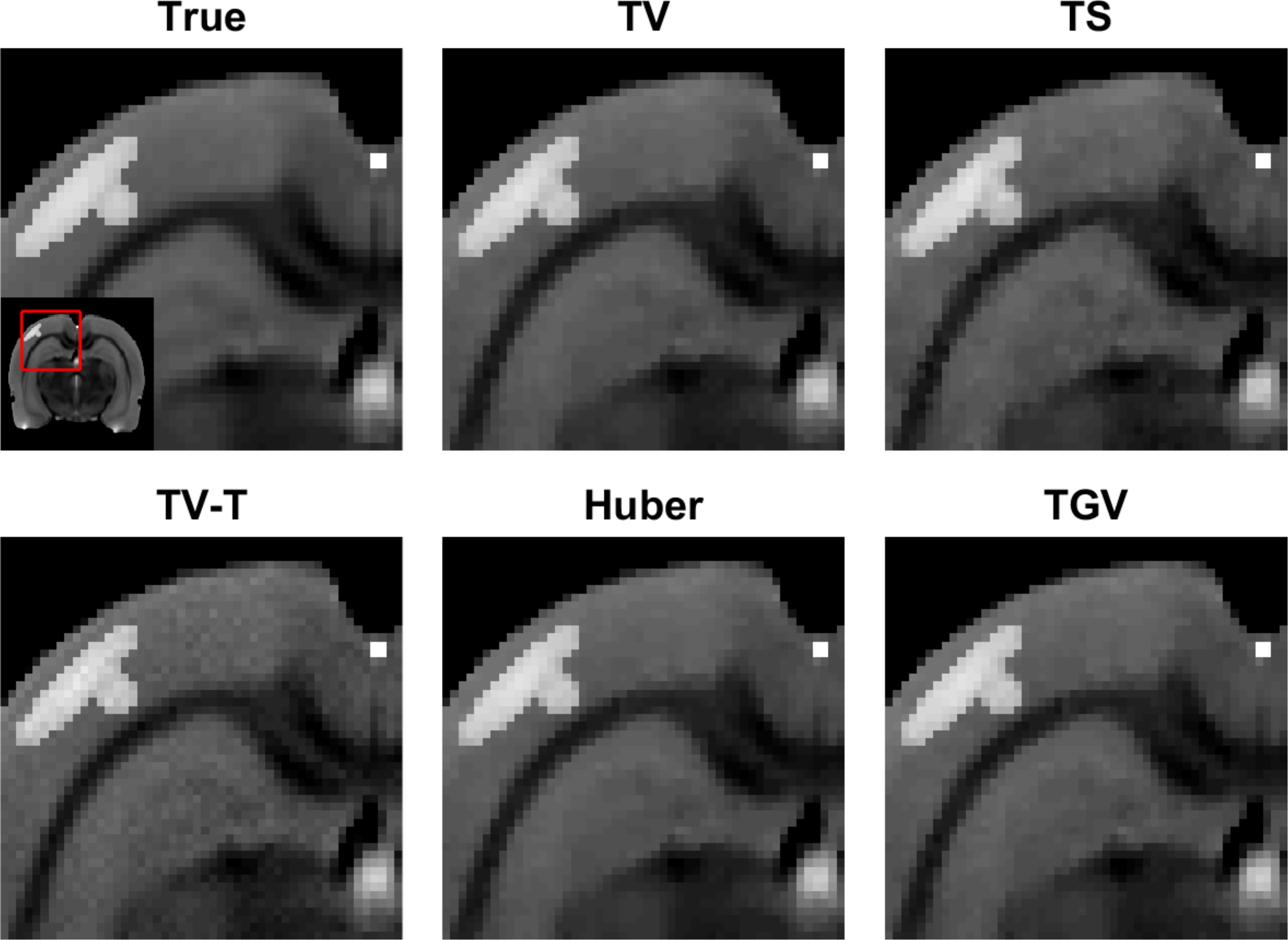}
    \caption{Closeups of the reconstructions at time t$\sim$100s with segment length of 34 showing the tumour ROI and the vascular ROI.}
    \label{fig:34_spatFrames}
\end{figure}

Fig.~\ref{fig:34_spatFrames} shows closeups of the reconstructions where both the tumour ROI and the vascular ROI are visible. Methods using spatial TV display mostly similar visual image quality. TV-T, which uses only temporal TV, shows visible deterioration in the single frame image quality as the image contains more spatial noise due to the lack of spatial regularization.

The number of iterations needed for the reconstructions, and the corresponding computation times are shown in Table \ref{tab:iter_times}. The computations were done in MATLAB (R2016b, The MathWorks, Inc., Natick, MA) on a server computer using 2 Intel Xeon E5-2630 v4 CPUs. The stopping criterion for the iterations was a relative change of less than $10^{-7}$ in the objective value in 10 consecutive iterations. The smooth temporal regularization methods, TS and Huber, exhibit faster convergence than the non-smooth methods.

Even though the computation times for a single iteration with the TV-T method were shorter than with the other methods, the total computation times were longer than with TV since the TV-T method required more iterations to reach convergence. With segment lengths 21 and 34, TV-T was the slowest to compute while TGV was the slowest with the other segment lengths.

\begin{figure}[hbtp]
	\centering
	\begin{subfigure}[b]{\columnwidth}
		\centering
		\includegraphics[width=.8\textwidth]{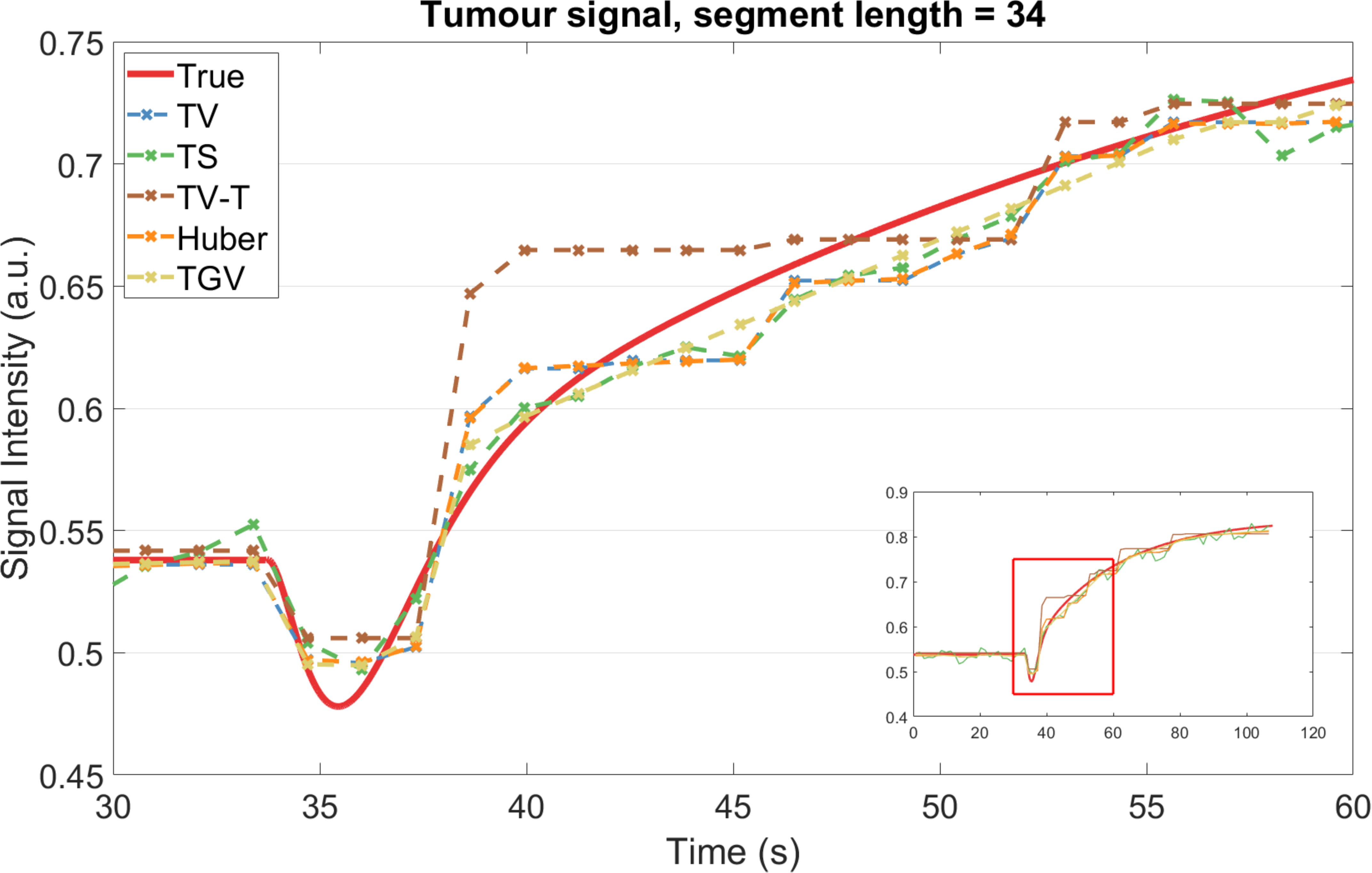}
	\end{subfigure}
	\newline
	\vspace{0cm}
	\begin{subfigure}[b]{\columnwidth}
		\centering
		\includegraphics[width=.8\textwidth]{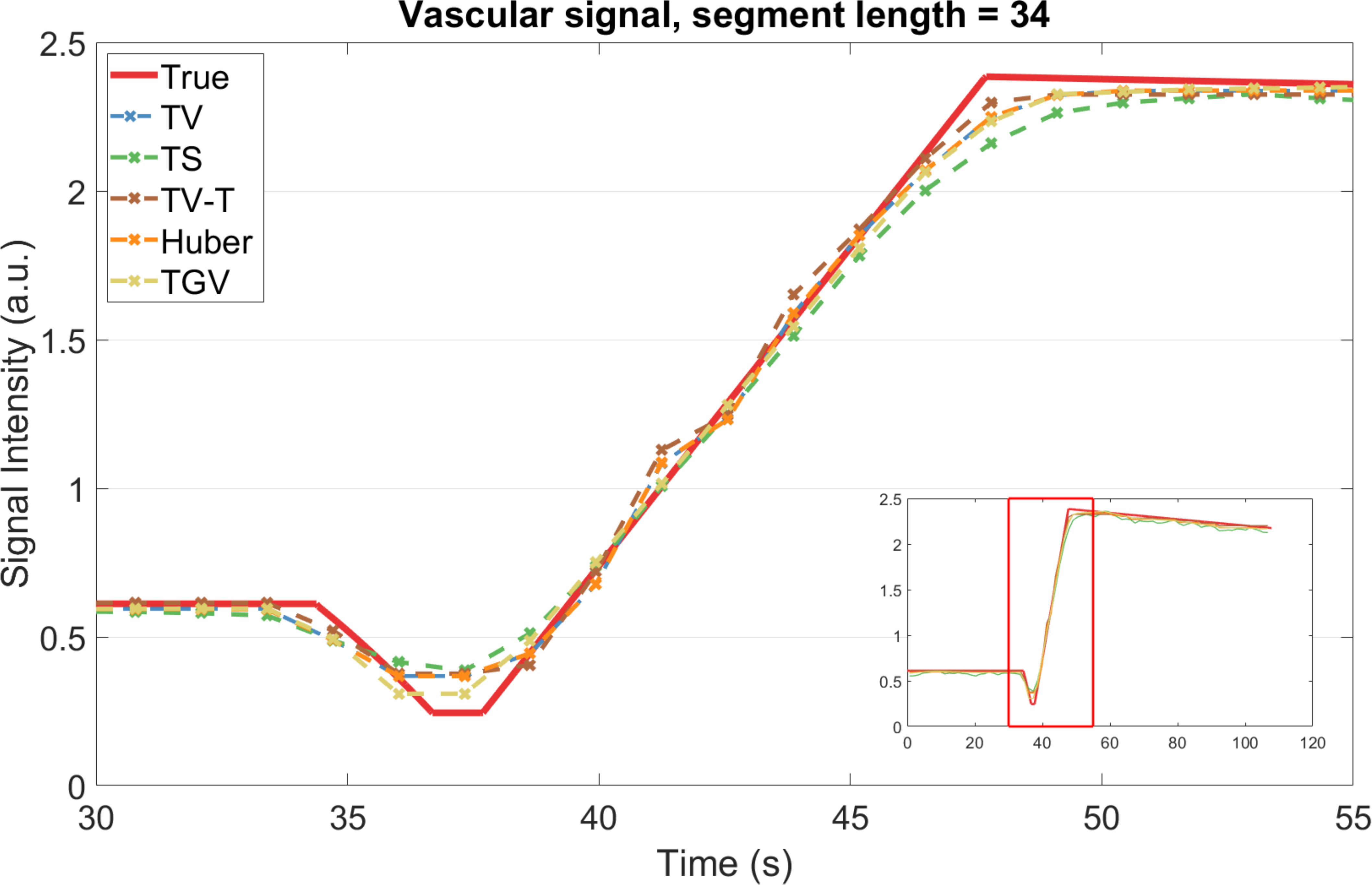}
	\end{subfigure}
	\caption{Single pixel signals from tumour (Top) and vascular (Bottom) regions at a segment length of 34 with the different methods at their optimal parameters according to the joint RMSE.}
	\label{fig:34_signals}
\end{figure}

Single pixel signals from the tumour and vascular regions are shown in Fig.~\ref{fig:34_signals}. In the tumour signal, the TV-T and TS reconstructions show more error to the true simulated signal. TGV has the best signal accuracy, while TV and Huber have similar signal accuracy with some staircasing visible. In the vascular signal, the TS reconstruction shows smoothing near both the minimum and the maximum of the signal. TGV has slightly better accuracy in the minimum of the vascular signal where TV, TV-T and Huber have similar reconstruction quality.

Huber reconstructions were also calculated for Huber parameters ranging from $\gamma=10^{-6}$ to $\gamma=0.1$. The joint RMSE of the reconstructions was very similar for the reconstructions with the parameter $\gamma$ ranging from $\gamma=10^{-6}$ to $\gamma=0.001$. For parameters larger than $\gamma=0.001$, the joint RMSE was closer to that of the TS model and thus the accuracy was worse. The computation times for the Huber reconstructions decreased when the Huber parameter was increased. Huber parameter $\gamma=0.001$ provided good balance between reconstruction accuracy and computation time.

\begin{table}[hbtp]
	\caption{Optimal temporal regularization parameters $\beta$ with different segment lengths and methods. The tested parameters were logarithmically even spaced.}
	\label{tab:opt_params}
	\footnotesize
	\centering
    \begin{tabular}{lllllll}
		\toprule
		& \multicolumn{6}{c}{Regularization parameter $\beta$}\\
		\cmidrule(lr){2-7}
		 & 8 & 13 & 21 & 34 & 55 & 89 \\
		\midrule
		TV & 0.1 & 0.056 & 0.01 & 0.01 & 0.0056 & 0.0056\\
		TS & 0.56 & 0.32 & 0.1 & 0.032 & 0.01 & 0.0001\\
		TV-T & 0.1 & 0.056 & 0.032 & 0.032 & 0.01 & 0.01\\
		Huber & 0.056 & 0.056 & 0.018 & 0.01 & 0.0056 & 0.0056\\
		TGV & 0.1 & 0.056 & 0.01 & 0.01 & 0.01 & 0.01\\
		\bottomrule
	\end{tabular}
\end{table}

\begin{table}[hbtp]
	\caption{Joint RMSEs with all segment lengths used and the five different methods. The best RMSE for each method is highlighted in red and the best RMSE for each segment length is bolded.}
	\label{tab:rmse_rois_res}
	\footnotesize
	\centering
	\begin{tabular}{lllllll}
		\toprule
		& \multicolumn{6}{c}{Joint RMSE}\\
		\cmidrule(lr){2-7}
		& 8 & 13 & 21 & 34 & 55 & 89 \\
		\midrule
		TV & 0.0425 & 0.0431 & 0.0413 & \color{red} 0.0379 & 0.0382 & 0.0513\\
		TS & 0.0555 & 0.0557 & 0.0553 & 0.0546 & \color{red} 0.0538 & 0.0595 \\
		TV-T & 0.0427 & 0.0416 & 0.0410 & \color{red} 0.0382 & 0.0404 & 0.0511 \\
		Huber & 0.0410 & 0.0408 & 0.0391 & 0.0374 & \color{red} 0.0373 & 0.0507\\
		TGV & \B 0.0366 & \B 0.0376 & \B 0.0373 & \B \color{red} 0.0354 & \B 0.0372 & \B 0.0505\\
		\bottomrule
	\end{tabular}
\end{table}

\begin{table}[hbtp]
	\caption{Computation times (top) and number of iterations (bottom) with the five different methods at different segment lengths.}
	\label{tab:iter_times}
	\footnotesize
	\centering
    \begin{tabular}{lllllll}
        \toprule
        & \multicolumn{6}{c}{Computation time (min) \& Iterations}\\
		\cmidrule(lr){2-7}
		& 8 & 13 & 21 & 34 & 55 & 89 \\
		\midrule
		TV & \myVertTab{203}{4600} & \myVertTab{107}{4047} & \myVertTab{14}{873} & \myVertTab{7.9}{757} & \myVertTab{4.2}{639} & \myVertTab{2.9}{621}	\vspace{0.1cm} \\
		TS & \myVertTab{9.4}{207} & \myVertTab{4.8}{179} & \myVertTab{2.8}{163} & \myVertTab{1.8}{176} & \myVertTab{1.2}{181} & \myVertTab{0.9}{197} \vspace{0.1cm} \\
		TV-T & \myVertTab{244}{7841} & \myVertTab{131}{6869} & \myVertTab{80}{6439} & \myVertTab{41}{5326} & \myVertTab{6.2}{1216} & \myVertTab{4.2}{1172} \vspace{0.1cm} \\
		Huber & \myVertTab{26}{550} & \myVertTab{14}{501} & \myVertTab{6.3}{352} & \myVertTab{2.2}{199} & \myVertTab{1.1}{160} & \myVertTab{0.7}{155} \vspace{0.1cm} \\
		TGV & \myVertTab{303}{6039} & \myVertTab{137}{4495} & \myVertTab{21}{1167} & \myVertTab{12}{1067} & \myVertTab{9.0}{1185} & \myVertTab{6.3}{1267} \\
		\bottomrule
	\end{tabular}
\end{table}

%%%%%%%%%%%%%%%%%%%%%%%%%%%%%%%%%%%%%%%%%%%%%%%%%%%%%%%%%%%%%%%%%
\subsection{In vivo measurements}

The in vivo measurements were reconstructed using the five different methods. Segment length was set to 34 for all reconstructions since this selection provided good reconstruction accuracy in the simulation. The temporal resolution of the image series was thus 1.309 s. The in-vivo measurements were scaled to be on the same intensity level as the simulation, so the same regularization parameters were used in the in vivo reconstructions as in the simulation reconstructions. Namely, these parameters were: $\alpha=0.001$ for all reconstructions, $\beta=0.01$ for TV, Huber and TGV, $\beta=0.032$ for TS and TV-T, $\gamma=0.001$ for Huber and $\gamma=\sqrt{2}$ for TGV, as is also shown in Table \ref{tab:opt_params} for the temporal regularization parameter $\beta$.

\begin{figure}[hbtp]
	\centering
	\includegraphics[width=.8\columnwidth]{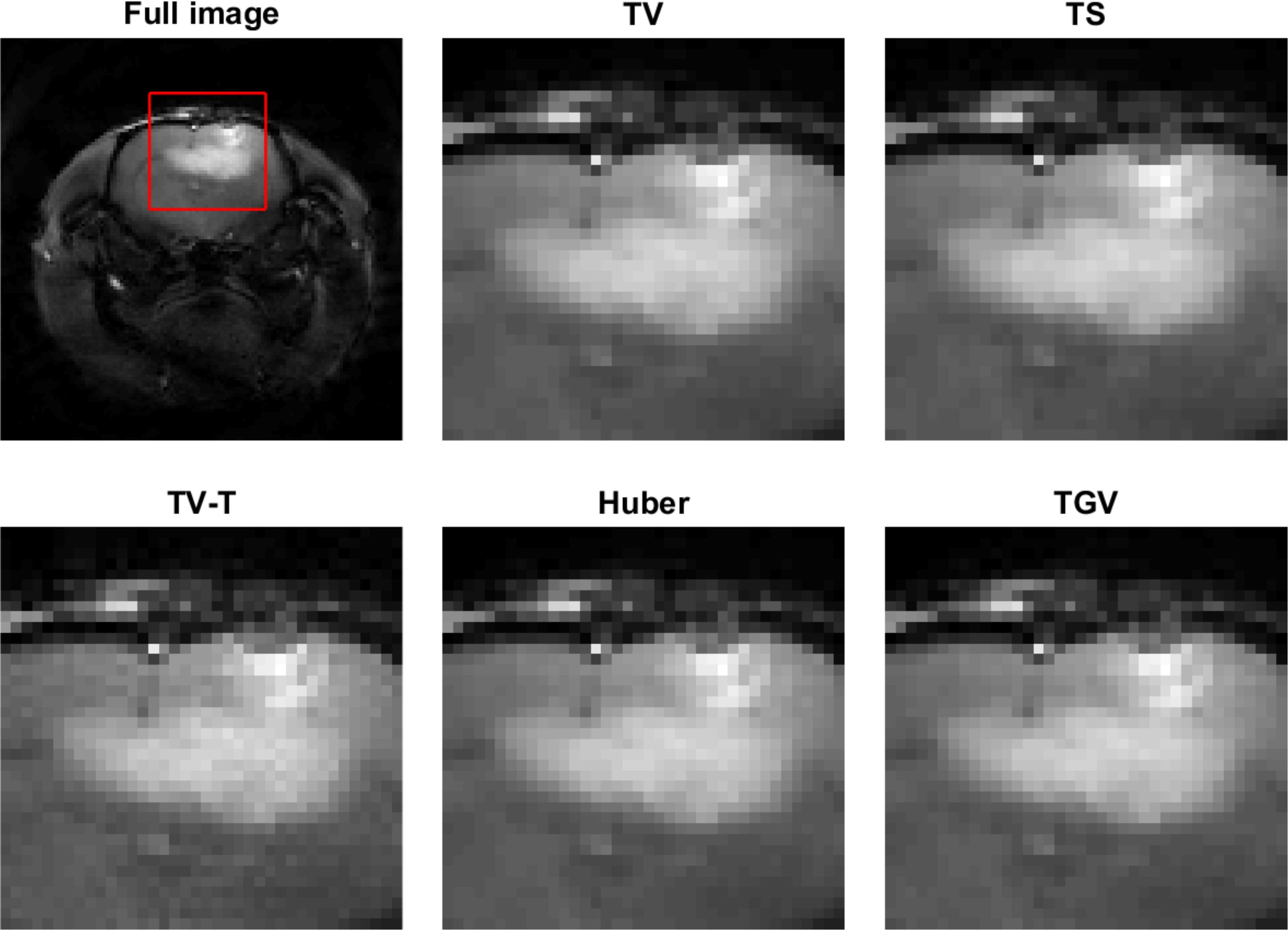}
	\caption{Closeups of the tumour area from the in vivo measurements with the five different methods approximately one minute after the injection of the contrast agent.}
	\label{fig:real_tumour}
\end{figure}

Fig.~\ref{fig:real_tumour} shows closeups of the tumour region of the reconstructions. Compared with the other methods, the TV-T model has worse visual image quality. All the methods using spatial regularization show visually similar image quality.

\begin{figure}[hbtp]
	\centering
	\includegraphics[width=.8\columnwidth]{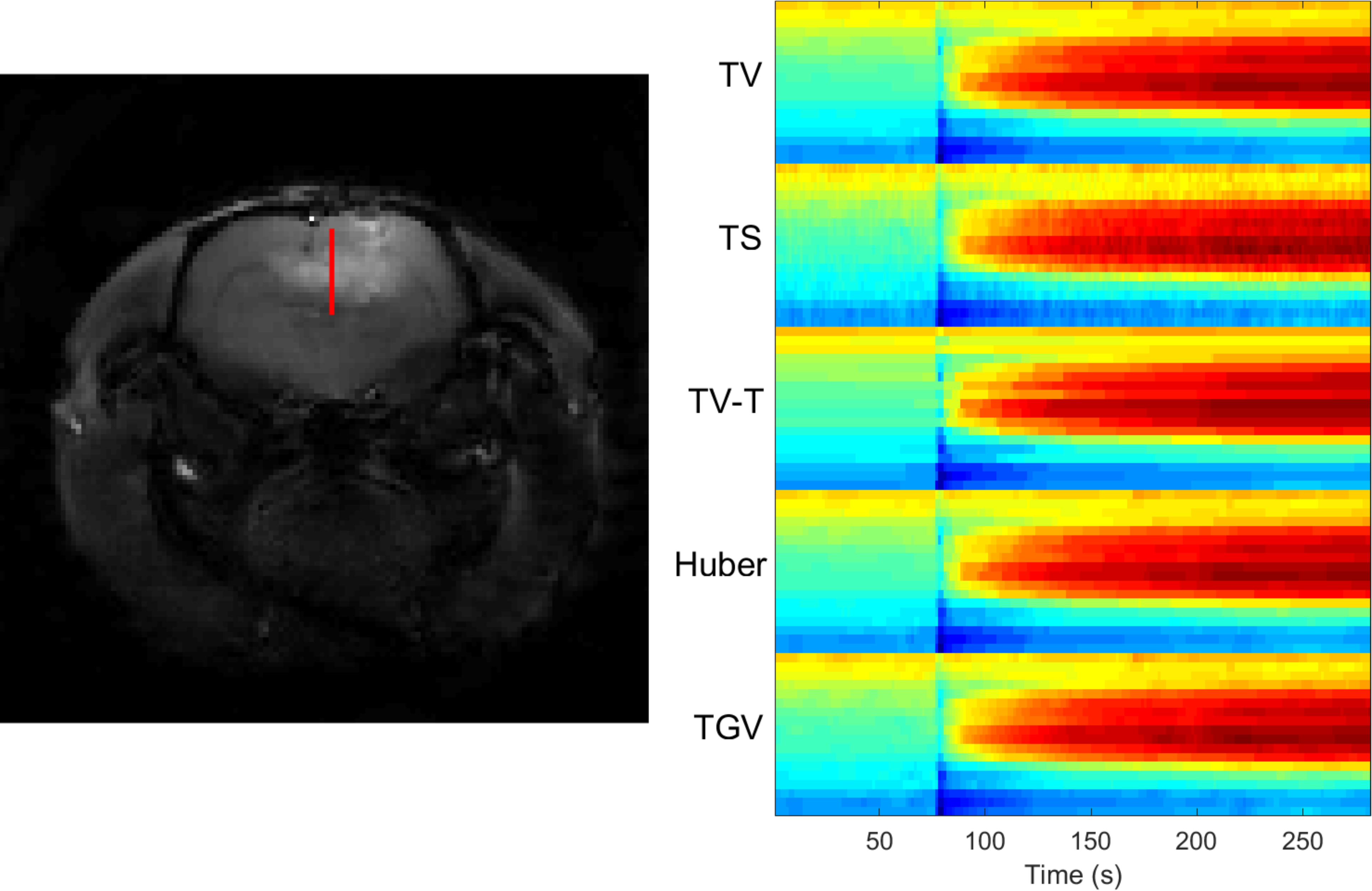}
	\caption{Time course of a line, which is marked in red in the left image, through the tumour in the in vivo dataset reconstructed with the different methods.}
	\label{fig:real_lineStack}
\end{figure}

Time courses of a line through the tumour area with the different reconstructions are shown in Fig.~\ref{fig:real_lineStack}. The TS reconstruction appears clearly noisy and the TV-T reconstruction has strong staircasing. The TV and Huber reconstructions suffer from some staircasing, whereas the TGV reconstruction is smooth and does not suffer from staircasing.

\begin{figure}[hbtp]
	\centering
	\begin{subfigure}[b]{\columnwidth}
		\centering
		\includegraphics[width=.8\textwidth]{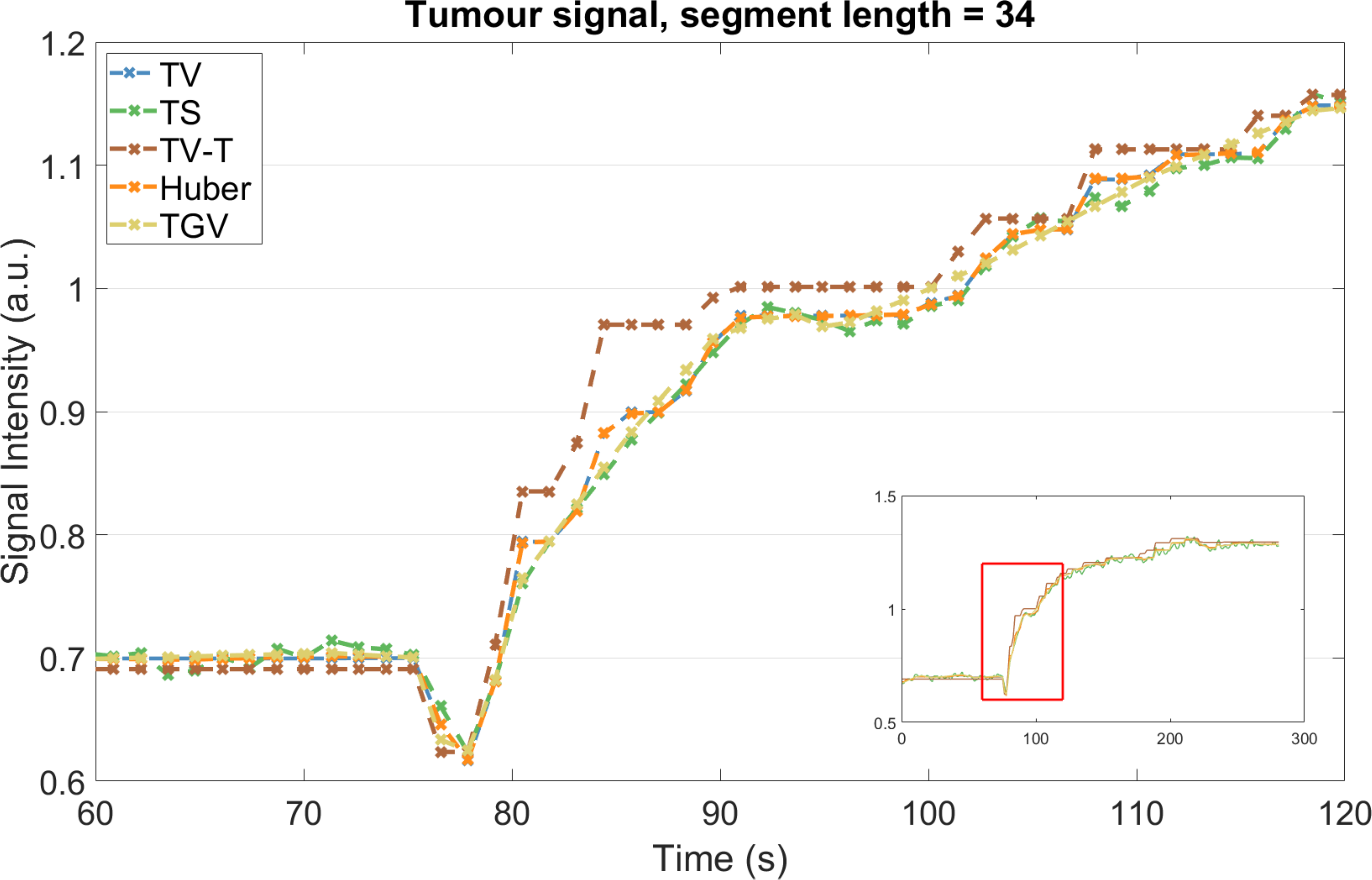}
	\end{subfigure}
	\newline
	\vspace{0cm}
	\begin{subfigure}[b]{\columnwidth}
		\centering
		\includegraphics[width=.8\textwidth]{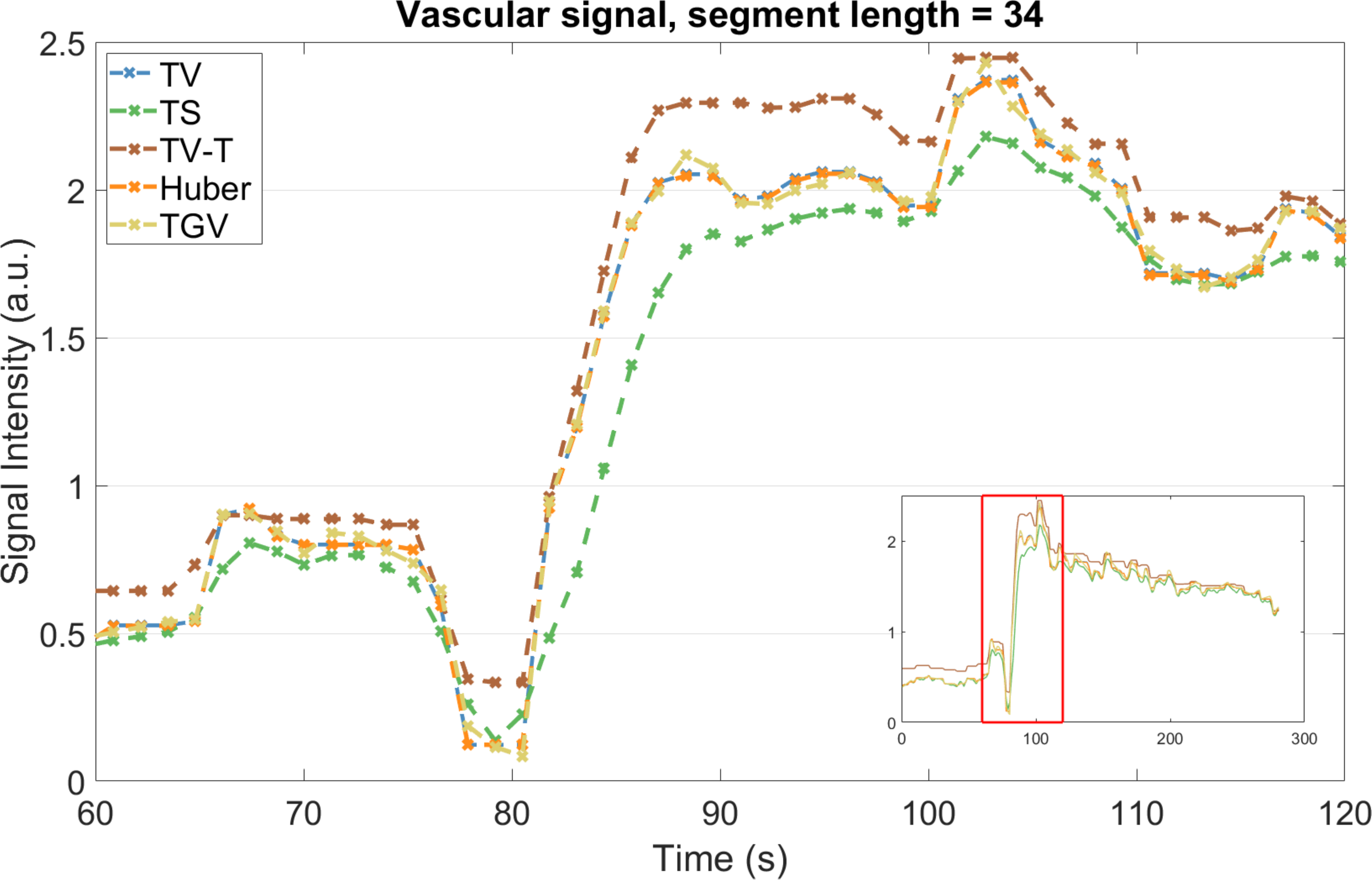}
	\end{subfigure}
	\caption{Single pixel signals from the reconstructions of the in vivo dataset from tumour (Top) and vascular (Bottom) regions. The images were reconstructed with a segment length of 34.}
	\label{fig:realMeas_signals}
\end{figure}

Fig.~\ref{fig:realMeas_signals} shows single pixel signals from the tumour and vascular regions with the different models. Here, the TV-T model exhibits clear staircasing effect, especially on the smooth tumour signal, and the TS model shows a smoothing effect in the sharp transient changes in the signal of the vascular region. The TV-T reconstruction shows higher intensity in the vascular signal than the other reconstructions due to the lack of spatial regularization. TV, Huber and TGV show mostly similar signal dynamics.

%%%%%%%%%%%%%%%%%%%%%%%%%%%%%%%%%%%%%%%%%%%%%%%%%%%%%%%%%%%%%%%%%
\section{Discussion}
In this work, we investigate the effects of four different temporal regularization models and six different segment lengths to reconstruction accuracy in DCE-MRI. The evaluations are carried out using both simulated and in vivo data. We also propose a new temporal regularization model; the Huber model.

The Huber model performs slightly better than the widely used TV in the simulation reconstruction accuracy with all segment lengths, and it is much faster to compute due to the smoothness of the temporal regularization functional. The Huber model is also quite insensitive to changes in the Huber parameter; changing the parameter from $\gamma=10^{-6}$ to $\gamma=0.001$ had almost no effect on the reconstruction accuracy and only affected the computational time.

TGV outperforms the other methods with all segment lengths in the simulation reconstruction accuracy measured by the joint RMSE. The method is able to reconstruct the smooth signal increase in the tumour area well while also being able to reconstruct the sharp transient signal changes in the vascular region accurately.

Using only temporal TV, the TV-T method performs well in the small vascular ROI reconstruction accuracy. The good accuracy in reconstructing the vascular ROI is likely due to the small size of the vascular ROI. The vascular ROI was only 4 pixels in size and therefore spatial regularization is likely to slightly dampen the signal variations in the ROI. In the larger tumour ROI and the rest of the image, the lack of spatial regularization results in worse reconstruction accuracy due to having stronger staircasing and higher spatial noise level. The results indicate that while temporal regularization is crucial for the high time resolution joint reconstruction of dynamic data, spatial regularization also significantly improves reconstruction quality.

Besides improving the reconstruction quality, spatial TV regularization also aided in convergence of the optimization problems. The TV-T reconstructions need\-ed the most iterations to converge with all segment lengths except 89 where TGV needed slightly more iterations to converge. The fastest method to compute was the TS method. The computation times of the Huber model with the used Huber parameter were much shorter than the computation times for the non-smooth total variation models. For TV, the computation times were approximately 2.2 to 7.8 times that of Huber and for TGV, they were 3.3 to 11.7 times that of Huber. The computation times of the Huber reconstructions depended on the Huber parameter; reconstructions with a small Huber parameter needed longer to converge. When looking at the reconstruction times, it should be noted that the ratio of the primal and dual step lengths affects the speed of convergence, and further optimization in the ratios could have an impact on the reconstruction times.

Using 34 spokes of measurements for each temporal frame outperformed the other segment lengths for all reconstruction methods except the TS and Huber methods, where a segment length of 55 was slightly better. Shorter segment lengths require stronger temporal regularization, which in turn leads to stronger staircasing for TV-based methods or peak diminishing for the TS model. Longer segment lengths in turn lack the temporal resolution to be able to accurately reconstruct sharp signal changes.

For TGV regularization the performance with short\-er segment lengths did not deteriorate as much as with the first order difference based methods. For the TS method, the differences in reconstruction accuracy with different segment lengths were also smaller, however, the accuracy was worse for all segment lengths.

The Huber model could also be modified to use a spatially varying Huber parameter, which would be smaller for regions with sharp signal changes such as in vascular regions, and higher for smoothly changing regions such as tumours. However, for a fair comparison this would need to be compared with spatially varying temporal regularization parameters for the other methods as well, which was out of the scope of this study.

In the simulation study we considered the combination of the golden angle and concentric squares sampling in order to obtain samples also from the corner areas of the k-space. The combination was demonstrated to lead to reduced aliasing artefacts in peripheral parts of the image domain when compared to the conventional radial sampling. In future studies, we seek to implement the scanning protocol for experimental studies.

%%%%%%%%%%%%%%%%%%%%%%%%%%%%%%%%%%%%%%%%%%%%%%%%%%%%%%%%%%%%%%%%%
\section{Conclusions}
In this paper, a new temporal regularization method, temporal Huber-regu\-larization, was proposed for DCE-MRI and the method was compared with three existing temporal regularization methods combined with spatial total variation regularization. The other methods were $L_2$-difference regularization (temporal smoothness, TS), $L_1$-difference regularization (temporal total variation, TV) and total generalized variation (TGV). The methods were also compared with temporal total variation without spatial regularization (TV-T). We found that Huber-regularization performs slightly worse than TGV, but outperforms the other methods. However, the computation times for Huber-regularization were reduced significantly compared to the TV and TGV methods, and especially large scale 4D DCE-MRI applications requiring fast computation could benefit significantly from using Huber model over TV or TGV.

All the methods were also tested for a wide range of segment lengths. In all cases a segment length of 34 provided good balance between reconstruction accuracy and computation time, and we expect that this gives a rough idea about a suitable segment length for joint reconstruction of golden angle DCE-MRI data of the brain. The best possible segment length, however, varies in practical applications, depending on the relation of the actual measurement speed (time per spoke) and the expected signal dynamics.

%%%%%%%%%%%%%%%%%%%%%%%%%%%%%%%%%%%%%%%%%%%%%%%%%%%%%%%%%%%%%%%%%
\section*{Acknowledgements}
This work was supported by the Academy of Finland (Project 312343, Finnish Centre of Excellence in Inverse Modelling and Imaging, 2018--2025) and the Jane and Aatos Erkko Foundation.

\printbibliography

\end{document}